\documentstyle[preprint,aps]{revtex}

\tightenlines

\begin{document}

\draft

\title{
Hyperon Non-leptonic Decays, the ${\bf |\Delta I|=1/2}$ Rule, and \\
A Priori Mixed Hadrons
}

\author{
A.~Garc\'{\i}a
}
\address{
Departamento de F\'{\i}sica.
Centro de Investigaci\'on y de Estudios Avanzados del IPN\\
A.P. 14-740, M\'exico, D.F., 07000, MEXICO \\ 
}
\author{ 
R.~Huerta
}
\address{
Departamento de F\'{\i}sica Aplicada \\
Centro de Investigaci\'on y de Estudios Avanzados del IPN. Unidad M\'erida \\ 
A.P. 73, Cordemex, M\'erida, Yucat\'an, 97310, MEXICO \\ 
}
\author{ 
G.~S\'anchez-Col\'on\cite{prmnnt}
}
\address{
Physics Department.
University of California \\
Riverside, California, 92521, U.S.A. \\
}

\date{March 11, 1998}

\maketitle

\begin{abstract}
The $|\Delta I|=1/2$ rule in non-leptonic decays of hyperons can be naturally
understood by postulating a priori mixed physical hadrons, along with the
isospin invariance of the responsible transition operator.
It is shown that this operator can be identified with the strong interaction
Yukawa hamiltonian.
The experimental amplitudes are well reproduced.
\end{abstract}


The possibility that strong-flavor and parity violating pieces in the mass 
operator of hadrons exist does not violate any known fundamental principle of 
physics.
If they do exist they would lead to non-perturbative a priori mixings 
of flavor and parity eigenstates in physical (mass eigenstates) hadrons.
Then, two paths for weak decays of hadrons to occur would be open: the ordinary
one mediated by $W^\pm_\mu$ ($Z_{\mu}$) and a new one via the strong-flavor and
parity conserving interaction hamiltonians.
The enhancement phenomenon observed in non-leptonic decays of hyperons (NLDH)
could then be attributed to this new mechanism.
However, for this to be the case it will be necessary that a priori mixings
produce the well established predictions of the $|\Delta I|=1/2$
rule\cite{marshak,donoghue}.

In this paper we shall (i) motivate the existence of a priori mixings, (ii) 
develop practical applications of such mixings via an ansatz which takes 
guidance in some model, (iii) show that indeed the predictions of the
$|\Delta I|=1/2$ in NLDH are obtained in this approach, and (iv) give a brief
account of the comparison of the amplitudes obtained with their experimental
values.

For motivation we shall use the model of Ref.\cite{barr}, in which the
electroweak sector is doubled along with the fermion and higgs content.
The gauge group is
$SU^C_3 \otimes SU_2\otimes U_1\otimes S\hat U_2 \otimes \hat U_1$,
there will be ordinary quarks $q$ and hatted (mirror) quarks $\hat q$ and two
doublet higgses $\phi$ and $\hat \phi$.
The latter will generate the mass matrix of the $q$'s and $\hat q$'s,
correspondingly.
After appropriate rotations the $q$'s and $\hat q$'s are assigned diagonal
masses and strong-flavors.
(See the diagonal terms in the matrix below.)
At this point we go beyond Ref.\cite{barr}: we assume the $q$'s and
$\hat q$'s to have opposite parities and that bisinglet and bidoublet higgsses
exist.
The diagonal mass matrix becomes (the calculation is straightforward; the
indices naught, $s$, and $p$ mean flavor, positive parity, and, negative parity
eigenstates, respectively, and we limit the discussion to $d$ and $s$ quarks;
the $u$, $c$, $b$, and $t$ quarks can be treated analogously)

\begin{equation}
\left(
\begin{array}{cccc}
{\bar d}_{0s} & {\bar s}_{0s} & {\bar d}_{0p} & {\bar s}_{0p}
\end{array}
\right)
\left(
\begin{array}{cccc}
m_{0d} & 0 & \Delta_{11} & \Delta_{12} \\
0 & m_{0s} & \Delta_{21} & \Delta_{22} \\
\Delta^*_{11} & \Delta^*_{21} & {\hat m}_{0d} & 0 \\
\Delta^*_{12} & \Delta^*_{22} & 0 & {\hat m}_{0s}  
\end{array}
\right)
\left(
\begin{array}{c}
d_{0s} \\ 
s_{0s} \\ 
d_{0p} \\ 
s_{0p}
\end{array}
\right)
\label{uno}
\end{equation}

A final rotation leads to the priori mixed physical (mass eigenstate) quarks, 
namely
$d_{ph} =
d_{0s} + \sigma s_{0s} + \delta s_{0p} + \cdots $, 
$s_{ph} =
s_{0s} - \sigma d_{0s} + \delta' d_{0p} + \cdots $, 
${\bar d}_{ph} =
{\bar d}_{0p} + \sigma {\bar s}_{0p} - \delta {\bar s}_{0s} + \cdots $, 
${\bar s}_{ph} =
{\bar s}_{0p} - \sigma {\bar d}_{0p} - \delta' {\bar d}_{0s} + \cdots $,
and similar expressions for $\hat d_{ph}$, etc.
Since necessarily
${\hat m}_{0d}$, ${\hat m}_{0s}$, ${\hat m}_{0s}-{\hat m}_{0d}\gg m_{0d}$,
$m_{0s}$, $\Delta_{ij}$, the angles in the last rotation can be kept to first
order.
There are six angles, three for $\Delta S = 0$ and three for $|\Delta S| = 1$
mixings.
The latter we have called $\sigma$, $\delta$, and $\delta'$. 
The dots stand for other mixings which will not be relevant in what follows. 
The above model shows how non-perturbative a priori mixings can arise.
An extended and more detailed discussion of the above approach is presented in
Refs.\cite{wrd}.

For practical applications of the above ideas one faces the problem of our 
current inability to compute well with QCD.
In order to proceed, one has no remedy but to develop an ansatz.
This latter will be based on the above model and it will consist of two steps:
(a) take the above mixings and (b) replace them in the non-relativistic quark
model (NRQM) wave functions.
This ansatz will yield a priori mixings at the hadron level.
We get at the meson level
$K^+_{ph} =
K^+_{0p} - \sigma \pi^+_{0p} - \delta' \pi^+_{0s} + \cdots $, 
$K^0_{ph} = 
K^0_{0p} +
\sigma \pi^0_{0p}/{\sqrt 2} +
\delta'\pi^0_{0s}/{\sqrt 2} + \cdots$, 
$\pi^+_{ph} = 
\pi^+_{0p} + \sigma K^+_{0p} - \delta K^+_{0s} + \cdots $, 
$\pi^0_{ph} =
\pi^0_{0p} -
\sigma ( K^0_{0p} + \bar K^0_{0p} )/{\sqrt 2} +
\delta ( K^0_{0s}- \bar K^0_{0s} )/{\sqrt 2} + \cdots $, 
$\pi^-_{ph} =
\pi^-_{0p} + \sigma K^-_{0p} + \delta K^-_{0s} + \cdots $, 
$\bar K^0_{ph} =
\bar K^0_{0p} + \sigma \pi^0_{0p}/{\sqrt 2} -
\delta '\pi^0_{0s}/{\sqrt 2} + \cdots $,
and
$K^-_{ph} =
K^-_{0p} - \sigma \pi^-_{0p} + \delta' \pi^-_{0s} + \cdots $. 
At the baryon level we get
$p_{ph} = 
p_{0s} - \sigma \Sigma^+_{0s} - \delta \Sigma^+_{0p} + \cdots $, 
$n_{ph} = 
n_{0s} + \sigma ( \Sigma^0_{0s}/{\sqrt 2} +
            \sqrt{3/2} \Lambda_{0s}) +
\delta ( \Sigma^0_{0p}/{\sqrt 2} +
            \sqrt{3/2} \Lambda_{0p} ) + \cdots $, 
$\Sigma^+_{ph} =
\Sigma^+_{0s} + \sigma p_{0s} - \delta' p_{0p} + \cdots $, 
$\Sigma^0_{ph} =
\Sigma^0_{0s} +
\sigma ( \Xi^0_{0s}- n_{0s} )/{\sqrt 2} +
\delta \Xi^0_{0p}/{\sqrt 2} + \delta' n_{0p}/{\sqrt 2} + \cdots $, 
$\Sigma^-_{ph} = \Sigma^-_{0s} + \sigma \Xi^-_{0s} + \delta \Xi^-_{0p} 
+ \cdots $,
$\Lambda_{ph} = 
\Lambda_{0s} + 
\sigma \sqrt{3/2} ( \Xi^0_{0s}- n_{0s}) +
\delta \sqrt{3/2} \Xi^0_{0p} + 
\delta' \sqrt{3/2} n_{0p} + \cdots $, 
$\Xi^0_{ph} =
\Xi^0_{0s} -
\sigma
( \Sigma^0_{0s}/{\sqrt 2} + \sqrt{3/2} \Lambda_{0s}) +
\delta'
( \Sigma^0_{0p}/{\sqrt 2} + \sqrt{3/2} \Lambda_{0p} ) + \cdots $,
and 
$\Xi^-_{ph} =
\Xi^-_{0s} - \sigma \Sigma^-_{0s} + \delta' \Sigma^-_{0p} + \cdots $.
Our phase conventions are those of Ref.\cite{lichtenberg}.
Notice that the physical mesons are $CP$-eigenstates, e.g.,
$CPK^+_{ph}=-K^-_{ph}$, etc., because we have assumed $CP$-invariance.

The a priori mixed hadrons will lead to NLDH via the parity and flavor
conserving strong interaction (Yukawa) hamiltonian $H_Y$.
The transition amplitudes will be given by the matrix elements
$\langle B_{ph}M_{ph}|H_Y|A_{ph}\rangle$, where $A_{ph}$ and $B_{ph}$ are the
initial and final hyperons and $M_{ph}$ is the emitted meson.
Using the above mixings these amplitudes will have the form
$\bar{u}_B(A-B\gamma_5)u_A$, where $u_A$ and $u_B$ are four-component Dirac
spinors and the amplitudes $A$ and $B$ correspond to the parity violating and
the parity conserving amplitudes of the $W^{\pm}_{\mu}$ mediated NLDH, although
with a priori mixings these amplitudes are both actually parity and flavor
conserving.
As a first approximation we shall neglect isospin violations, i.e., we shall
assume that $H_Y$ is an $SU_2$ scalar.
However, we shall not neglect $SU_3$ breaking.
One obtains for $A$ and $B$ the results:

\[
A_1
=
\delta'
\sqrt 3 g^{{}^{p,sp}}_{{}_{p,p\pi^0}} +
\delta
(
g^{{}^{s,ss}}_{{}_{\Lambda,pK^-}} - g^{{}^{s,pp}}_{{}_{\Lambda,\Sigma^+\pi^-}}
)
,\ \ \ 
A_2
=
-
[
\delta'
\sqrt 3 g^{{}^{p,sp}}_{{}_{p,p\pi^0}} +
\delta
(
g^{{}^{s,ss}}_{{}_{\Lambda,pK^-}} - g^{{}^{s,pp}}_{{}_{\Lambda,\Sigma^+\pi^-}}
)
]/{\sqrt 2} 
,\]

\begin{equation}
A_3
=
\delta
(
\sqrt 2 g^{{}^{s,ss}}_{{}_{\Sigma^0,p K^-}} +
\sqrt{3/2} g^{{}^{s,pp}}_{{}_{\Sigma^+,\Lambda\pi^+}} +
g^{{}^{s,pp}}_{{}_{\Sigma^+,\Sigma^+\pi^0}}/{\sqrt 2}
)
,
\label{cuatro}
\end{equation}

\[
A_4
=
-\delta'
\sqrt 2 g^{{}^{p,sp}}_{{}_{p,p\pi^0}} +
\delta
(
\sqrt{3/2} g^{{}^{s,pp}}_{{}_{\Sigma^+,\Lambda\pi^+}} -
g^{{}^{s,pp}}_{{}_{\Sigma^+,\Sigma^+\pi^0}}/{\sqrt 2}
)
,\ \ \ 
A_5
=
-\delta'
g^{{}^{p,sp}}_{{}_{p,p\pi^0}} -
\delta
(
g^{{}^{s,ss}}_{{}_{\Sigma^0,pK^-}} +
g^{{}^{s,pp}}_{{}_{\Sigma^+,\Sigma^+\pi^0}}
)
,
\]

\[
A_6
=
\delta'
g^{{}^{p,sp}}_{{}_{\Sigma^+,\Lambda\pi^+}} +
\delta
(
g^{{}^{s,ss}}_{{}_{\Xi^-,\Lambda K^-}} + 
\sqrt 3 g^{{}^{s,pp}}_{{}_{\Xi^0,\Xi^0\pi^0}}
)
,\ \ \ 
A_7
=
[
\delta'
g^{{}^{p,sp}}_{{}_{\Sigma^+,\Lambda\pi^+}} +
\delta
(
g^{{}^{s,ss}}_{{}_{\Xi^-,\Lambda K^-}} + 
\sqrt 3 g^{{}^{s,pp}}_{{}_{\Xi^0,\Xi^0\pi^0}}
)
]/{\sqrt 2}
,\]

\noindent
and

\[
B_1
=
\sigma
(
- \sqrt 3 g_{{}_{p,p\pi^0}} +
g_{{}_{\Lambda,pK^-}} - g_{{}_{\Lambda,\Sigma^+\pi^-}}
)
,\ \ \ 
B_2
=
-
\sigma
(
- \sqrt 3 g_{{}_{p,p\pi^0}} +
g_{{}_{\Lambda,pK^-}} - g_{{}_{\Lambda,\Sigma^+\pi^-}}
)/{\sqrt 2}
,\]

\begin{equation}
B_3
=
\sigma
(
\sqrt 2 g_{{}_{\Sigma^0,p K^-}} +
\sqrt{3/2} g_{{}_{\Sigma^+,\Lambda\pi^+}} +
g_{{}_{\Sigma^+,\Sigma^+\pi^0}}/{\sqrt 2}
)
,
\label{cinco}
\end{equation}

\[
B_4
=
\sigma
(
\sqrt 2 g_{{}_{p,p\pi^0}} +
\sqrt{3/2} g_{{}_{\Sigma^+,\Lambda\pi^+}} -
g_{{}_{\Sigma^+,\Sigma^+\pi^0}}/{\sqrt 2}
)
,\ \ \ 
B_5
=
\sigma
(
g_{{}_{p,p\pi^0}} -
g_{{}_{\Sigma^0,pK^-}} - g_{{}_{\Sigma^+,\Sigma^+\pi^0}}
)
,
\]

\[
B_6
=
\sigma
(
- g_{{}_{\Sigma^+,\Lambda\pi^+}} +
g_{{}_{\Xi^-,\Lambda K^-}} + 
\sqrt 3 g_{{}_{\Xi^0,\Xi^0\pi^0}}
)
,\ \ \ 
B_7
=
\sigma
(
- g_{{}_{\Sigma^+,\Lambda\pi^+}} +
g_{{}_{\Xi^-,\Lambda K^-}} + 
\sqrt 3 g_{{}_{\Xi^0,\Xi^0\pi^0}}
)/{\sqrt 2}
.\]

\noindent
The subindeces $1, \dots, 7$ correspond to
$\Lambda\rightarrow p\pi^-$,
$\Lambda\rightarrow n\pi^0$, 
$\Sigma^-\rightarrow n\pi^-$, 
$\Sigma^+\rightarrow n\pi^+$, 
$\Sigma^+\rightarrow p\pi^0$, 
$\Xi^-\rightarrow \Lambda\pi^-$,
and 
$\Xi^0\rightarrow \Lambda\pi^0$,
respectively.
The $g$-constants in these equations are Yukawa coupling constants (YCC)
defined by the matrix elements of $H_Y$ between flavor and parity eigenstates,
for example, by
$\langle B_{0s} M_{0p} |H_Y|A_{0p}\rangle=g^{{}^{p,sp}}_{{}_{A,BM}}$.
We have omitted the upper indeces in the $g$'s of the $B$ amplitudes because
the states involved carry the normal intrinsic parities of hadrons.
In Eqs.~(\ref{cinco}) we have used the $SU_2$ relations
$g_{{}_{p,p\pi^0}}=-g_{{}_{n,n\pi^0}}=g_{{}_{p,n\pi^+}}/{\sqrt 2}
=g_{{}_{n,p\pi^-}}/{\sqrt 2}$,
$g_{{}_{\Sigma^+,\Lambda\pi^+}}=g_{{}_{\Sigma^0,\Lambda\pi^0}}
=g_{{}_{\Sigma^-,\Lambda\pi^-}}$,
$g_{{}_{\Lambda,\Sigma^+\pi^-}}=g_{{}_{\Lambda,\Sigma^0\pi^0}}$,
$g_{{}_{\Sigma^+,\Sigma^+\pi^0}}=-g_{{}_{\Sigma^+,\Sigma^0\pi^+}}
=g_{{}_{\Sigma^-,\Sigma^0\pi^-}}$,
$g_{{}_{\Sigma^0,pK^-}}=g_{{}_{\Sigma^-,nK^-}}/{\sqrt 2}
=g_{{}_{\Sigma^+,p\bar K^0}}/{\sqrt 2}$,
$g_{{}_{\Lambda,pK^-}}=g_{{}_{\Lambda,n\bar K^0}}$,
$g_{{}_{\Xi^0,\Xi^0\pi^0}}=g_{{}_{\Xi^-,\Xi^0\pi^-}}/{\sqrt 2}$,
$g_{{}_{\Xi^-,\Lambda K^-}}=-g_{{}_{\Xi^0,\Lambda \bar K^0}}$,
and
$g_{{}_{\Lambda,\Lambda \pi^0}}=0$.
Similar relations are valid within each set of upper indeces, e.g.,
$g^{{}^{p,sp}}_{{}_{p,p\pi^0}}=-g^{{}^{p,sp}}_{{}_{n,n\pi^0}}$, etc.;
the reason for this is, as we discussed in Ref.~\cite{wrd}, mirror hadrons may
be expected to have the same strong-flavor assignments as ordinary hadrons.
Thus, for example, $\pi^+_{0s} $, $\pi^0_{0s} $, and $\pi^-_{0s} $ form an
isospin triplet, although a diferent one from the ordinary $\pi^+_{0p} $,
$\pi^0_{0p} $, and $\pi^-_{0p} $ isospin triplet.
These latter relations have been used in Eqs.~(\ref{cuatro}).

From the above results one readily obtains the equalities:

\begin{equation}
A_2 = 
-A_1/{\sqrt 2},\ \ \ \ \ \ 
A_5 =
(
A_4-A_3
)/{\sqrt 2},\ \ \ \ \ \ 
A_7 = 
A_6/{\sqrt 2},
\label{seis}
\end{equation}

\begin{equation}
B_2 = 
-B_1/{\sqrt 2},\ \ \ \ \ \
B_5 =
(
B_4-B_3
)/{\sqrt 2},\ \ \ \ \ \ 
B_7 = 
B_6/{\sqrt 2}.
\label{siete}
\end{equation}

\noindent
These are the predictions of the $|\Delta I|=1/2$ rule.
That is, a priori mixings in hadrons as introduced above lead to the
predictions of the $|\Delta I|=1/2$ rule, but notice that they do not lead to
the $|\Delta I|=1/2$ rule itself.
This rule originally refers to the isospin covariance properties of the
effective non-leptonic interaction hamiltonian to be sandwiched between
strong-flavor and parity eigenstates.
The $I=1/2$ part of this hamiltonian is enhanced over the $I=3/2$ part.
In contrast, in the case of a priori mixings $H_Y$ has been assumed to be
isospin invariant, i.e., in this case the rule should be called a
$\Delta I=0$ rule.

It must be stressed that the results~(\ref{seis}) and (\ref{siete}) are very
general: (i) the predictions of the $|\Delta I|=1/2$ rule are obtained
simultaneously for the $A$ and $B$ amplitudes, (ii) they are independent of
the mixing angles $\sigma$, $\delta$, and $\delta'$, and (iii) they are also
independent of particular values of the YCC.
They will be violated by isospin breaking corrections.
So, they should be quite accurate, as is experimentally the case.

Although a priori mixings do not violate any fundamental principle, the reader
may wonder if they do not violate some important theorem, specifically the 
Feinberg--Kabir--Weinberg theorem\cite{feinberg}.
They do not.
This theorem  is useful for defining conserved quantum numbers after rotations
that diagonalize the kinetic and mass terms of particles.
It presupposes on mass-shell particles and interactions that can be
diagonalized simultaneously with those terms.
This last is sometimes not clearly stated, but it is an obvious requirement.
Quarks inside hadrons are off mass-shell; so the theorem cannot eliminate the
non-diagonal $d$-$s$ terms which lead to non-diagonal terms in hadrons.
It has not yet been proved for hadrons, but one can speculate: what if it had?
Hadrons are on mass-shell, but they show many more interactions than quarks,
albeit, effective ones.
The Yukawa interaction cannot be diagonalized along with the kinetic and mass
terms, as can be seen through the YCC of the amplitudes above.
Therefore, this theorem would not apply to the last rotation leading to 
a priori mixings in hadrons.
Another example is weak radiative decays, it is interesting because it is a
mixed one.
The charge form factors can be diagonalized while anomalous magnetic ones
cannot.
The theorem would apply to the former but not to the latter.

The reader may wonder where specifically the predictions of the $|\Delta I| = 
1/2$ came from.
They can be traced down to the coefficients of $\sigma$, $\delta$, and
$\delta'$ in the mixed hadrons,
$1/\sqrt{2}$, $\sqrt{3/2}$, etc., and the latter in turn came from
reconstructing the NRQM wave function.
In this respect, there is an important comment we wish to make.
The factorization of these coefficients and the angles from the NRQM wave
functions should be preserved by QCD, because QCD did not entervene at all in
their fixing and it treats all quarks on an equal footing.
In other words, the effect forming compound hadrons by setting the quarks in
motion and in interaction with one another will go into rendering the NRQM wave
functions into realistic strong-flavor and parity eigenstate wave functions,
but should not break the above factorization.
One may expect Eqs.~(\ref{seis}) and (\ref{siete}) to remain correct after QCD
fully operates.
The important question is whether one has results that are valid beyond the
particular models one has taken for guidance.
This argument supports the affirmative answer.

A detailed comparison with all the experimental data available in these decays
requires more space and will be presented separately~\cite{nll}.
Nevertheless, we shall briefly mention a few very important results.
First, the experimental $B$ amplitudes~\cite{pdg} (displayed in
Table~\ref{table1})
are reproduced within a few percent by accepting that the YCC are given by the
ones observed in strong interactions~\cite{dumbrajs}, an assumption which
cannot be avoided in this approach.
The best predictions for these amplitudes are 
$B_1=22.11\times 10^{-7}$,
$B_2=-15.63\times 10^{-7}$,
$B_3=1.39\times 10^{-7}$,
$B_4=-42.03\times 10^{-7}$,
$B_5=-30.67\times 10^{-7}$,
$B_6=17.45\times 10^{-7}$,
and 
$B_7=12.34\times 10^{-7}$. 
The only unknown parameter $\sigma$ is determined at 
$(3.9\pm 1.3)\times 10^{-6}$. 
Second, although the $A$ amplitudes involve new YCC, an important prediction is
already made in Eqs.~(\ref{cuatro}).
Once the signs of the $B$ amplitudes are fixed, one is free to fix the signs of
four $A$ amplitudes --- say, $A_1>0$, $A_3<0$, $A_4<0$, $A_6<0$ --- to match
the signs of the corresponding experimental $\alpha$ asymmetries, namely,
$\alpha_1>0$, $\alpha_3<0$, $\alpha_4>0$, $\alpha_6<0$~\cite{pdg}.
Then the signs of $A_2<0$, $A_5>0$, and $A_7<0$ are fixed by
Eqs.~(\ref{cuatro}) and the fact that $|A_4|\ll |A_3|$.
In turn the signs of the corresponding $\alpha$'s are fixed.
These three signs agree with the experimentally observed ones, namely,
$\alpha_2>0$, $\alpha_5<0$, $\alpha_7<0$.

The above predictions are quite general because only assumptions already
implied in the ansatz for the application of a priori mixings have been used.
A detailed comparision of the $A$ amplitudes with experiment is limited by our
current inability to compute well with QCD.
However, one may try simple and argumentable new assumptions to make
predictions for such amplitudes.
Since QCD has been assumed to be common to both ordinary and mirror quarks, it
is not unreasonable to expect that the magnitudes of the YCC in the $A$
amplitudes have the same magnitudes as their corresponding counterparts in the
ordinary YCC of the $B$ amplitudes.
The relative signs may differ, however.
Introducing this assumption we obtain the predictions for the $A$ amplitudes
displayed in Table~\ref{table1}.
The predictions for the $B$ amplitudes must also be redone, because
determining the $A$ amplitudes alone may introduce small variations in the YCC
that affect importantly the $B$ amplitudes, i.e., both the $A$ and $B$
amplitudes must be simultaneously determined, the $B$'s act then as extra
constraints on the determination of the $A$'s.
The new predictions for the $B$'s are also displayed in Table~\ref{table1}.
In obtaining Table~\ref{table1} we have actually used the experimental decay
rates $\Gamma$ and $\alpha$ and $\gamma$ asymmetries, but we only display the
experimental and theoretical amplitudes.

The predictions for the $A$'s agree very well with experiment to within a few
percent, while the predictions for the $B$'s remain as before.
The a priori mixing angles are determined to be
$\delta=(0.22\pm 0.04)\times 10^{-6}$,
$\delta'=(0.25\pm 0.04)\times 10^{-6}$,
and
$\sigma=(4.6\pm 0.8)\times 10^{-6}$.
This last value of $\sigma$ is consistent with the previous one.
The more detailed analysis of the comparison of the $A$'s and $B$'s with
experiment is presented in Ref.~\cite{nll}.

The above results, especially those of Eqs.~(\ref{seis}) and (\ref{siete}) and
the determination of the amplitudes, satisfy some of the most important
requirements that a priori mixings must meet in order to be taken seriously as
an alternative to describe the enhancement phenomenon observed in non-leptonic
decays of hadrons.
This means then that another source of flavor and parity violation may exist,
other than that of $W^{\pm}_{\mu}$ and $Z_{\mu}$.
It is worthwhile to point out that the calculation of decays and reactions
through the $W/Z$ exchange mechanisms is obtained in the present scheme in the
usual way.
The weak hamiltonian is, so to speak, sandwiched between a priori mixed
hadrons; to lowest order only the parity and flavor eigenstates
survive, the mixed eigenstates contribute negligible corrections. 
Thus, beta and semileptonic decay remain practically unchanged,
while nonleptonic kaon decays, hypernuclear decays, and others in
which the enhancement phenomenon could be present should be recalculated.

We would like to thank CONACyT (M\'exico) for partial support.

\begin{table}
\caption{
Predictions for the $A$ amplitudes, along with the accompanying predictions for
the $B$ amplitudes, obtained by assuming that the magnitudes of the YCC of
Eqs.~(\ref{cuatro}) match their corresponding counterparts in
Eqs.~(\ref{cinco}).
The values of the YCC are listed in Ref.~[9].
All amplitudes are given in units of $10^{-7}$.
}~
\label{table1}
\begin{tabular}
{
r@{$\rightarrow$}l
r@{.}l@{\,$\pm$\,}r@{.}l
d
r@{.}l@{\,$\pm$\,}r@{.}l
d
}
\multicolumn{2}{c}{Decay} &
\multicolumn{4}{c}{$B_{\rm exp}$} &
$B_{\rm th}$ &
\multicolumn{4}{c}{$A_{\rm exp}$} &
$A_{\rm th}$
\\
\tableline
$\Lambda$ & $p\pi^-$ &
$-$22 & 09 & 0 & 44 &
$-$22.36 & 
$-$3 & 231 & 0 & 020 &
$-$3.263
\\
$\Lambda$& $n\pi^0$ &
15 & 89 & 1 & 01 &
15.81 & 
2 & 374 & 0 & 027 &
2.308
\\
$\Sigma^-$ & $n\pi^-$ &
1 & 43 & 0 & 17 &
1.35 & 
$-$4 & 269 & 0 & 014 &
$-$4.264
\\
$\Sigma^+$ & $n\pi^+$ &
$-$42 & 17 & 0 & 18 &
$-$42.10 & 
$-$0 & 140 & 0 & 027 &
$-$0.153
\\
$\Sigma^+$ & $p\pi^0$ &
$-$26 & \multicolumn{3}{l}{$86{\ }^{+\ 1.10}_{-\ 1.36}$} &
$-$30.72 & 
3 & \multicolumn{3}{l}{$247{\ }^{+\ 0.089}_{-\ 0.116}$} &
2.907
\\
$\Xi^-$ & $\Lambda\pi^-$ &
$-$17 & 47 & 0 & 50 &
$-$17.28 & 
4 & 497 & 0 & 020 &
4.521
\\
$\Xi^0$ & $\Lambda\pi^0$ &
$-$12 & 29 & 0 & 70 &
$-$12.22 & 
3 & 431 & 0 & 055 &
3.197
\\
\end{tabular}
\end{table}


\begin{references}
\bibitem{pdg}
Particle Data Group, Phys.\ Rev.\ {\bf D 50}, 1173 (1994).
We quote the experimental values of the $B$ amplitudes in the natural scale of
$10^{-7}$, see Ref.~\cite{donoghue}.
Their signs are free to choose; actually, the comparison with theoretical
predictions is only meaningful for their magnitudes.
The signs we display in the text are for convenience only.
This is not the case for the signs in the $A$ amplitudes, see later in the text.
\end{references}
\end{document}